\documentclass[12pt]{article}
\usepackage[dvips]{graphicx}
\begin{document}
\bibliographystyle{prsty}
\pagestyle{empty}
\begin{flushright}
{CERN-PH-TH/2006-117\\
math-ph/0510065 v5}
\end{flushright}
\vspace*{10mm}
\begin{center}
{\bf ON THE STABILITY OF FOUR-LEGGED TABLES} \\
\vspace*{1cm} 
{\bf A. Martin} \\
\vspace{0.3cm}
CERN, Department of Physics, TH Division \\
CH - 1211 Geneva 23 \\
\vspace*{3cm}  
{\bf ABSTRACT} \\ \end{center}
\vskip 1cm
We prove that a perfect square table with four legs, placed on continuous irregular ground with a local slope of at most 14.4 degrees and later 35 degrees, can be put into equilibrium on the ground by a ``rotation" of less than 90 degrees.  We also discuss the case of non-square tables and make the conjecture that equilibrium can be found if the four feet lie on a circle.

\vspace*{4.0cm}

\begin{flushleft} CERN-PH-TH/2006-117\\math-ph/0510065 v5 \\
August 2006
\end{flushleft}
\vfill\eject

\setcounter{page}{1}
\pagestyle{plain}

\section{
Introduction and History}

Many people eating lunch or drinking coffee on the terrace of the CERN cafeteria have had the following problem: the table is often not in a stable equilibrium position. It rests on three feet and with very little energy it can be made to wobble so that part of your coffee is spilled, at best onto the saucer or at worst onto the table. Why is this? Not because the table is not well built but because the ground is very irregular. Many years ago I thought about this problem, and in a relatively idealized situation I proved that by ``rotating" the table (``rotating" is to be explained below) one could find an equilibrium position if the local slope is less than 15 degrees. I carried out the experiment many times on the terrace, and even though the conditions of the theorem are not really satisfied - the feet are thick, the ground is sometimes discontinuous, but on the other hand, the legs of the tables have  some elasticity - I have always succeeded in finding an equilibrium position. I think that my early sketch of a proof dates back to 1995.  I presented a  tentatively ``serious" (but partly incorrect) proof in 1998 at a seminar at the Institut des Hautes \'{E}tudes Scientifiques (IH\'{E}S) in Bures-sur-Yvette (a witness is Professor Marcel Berger, former director of IHES)  but it was never written up due to serious personal reasons.

Perhaps I should apologize for using the word ``stability" in the title, because this is not stability in the mathematical sense, but the fact that one tries to find a position where the table does not wobble. 

On 1 October 2005, a first version of the present work was submitted to the Archives of Mathematical Physics and accepted by the Archives \cite{ref1a}.  It was proved that it was possible to find a position where the 4 feet of the square table were resting on the ground.  We reproduce this proof here.  This condition was that the local slope should be less than 14.47 degrees.  It was also conjectured that it was possible to put the 4 feet of a table on ground provided the 4 feet were on a circle.

On 19 November, a paper on the same subject appeared in the Archives of mathematics \cite{ref2a}.  In this paper a different approach was used leading to the condition that the local slope should be less than 35.26 degrees.  Their proof applied not only to square tables but also to rectangular tables (notice that this is a special case of a table with 4 feet on a circle!).  They also give an extensive list of references, including Ref. \cite{ref1a}.  Among these, I wish to single out the paper by Dyson \cite{ref3a} which presents a result both stronger and weaker than the one presented in Refs. \cite{ref1a} and \cite{ref2a}.  He shows that on a sphere, there are four points on two diameters at right angle where a function, defined on the sphere takes the same value. One can think of the function as the height of mountains.  The result is stronger because the heights are equal while our four-legged table could  have 4 feet on the ground but the table could be inclined.   On the other hand, we do not have to consider the surface globally.  It is sufficient, both in \cite{ref1a} and \cite{ref2a} to have a control on a finite region of the surface, of the order of 2--3 times  the dimensions of the table in length.  Further results, following Dyson are quoted in Ref.  \cite{ref2a}, like Ref. \cite{ref4a}.  One can also single out qualitative arguments \cite{ref5a}\cite{ref5b},\cite{ref6a} which have the merit of being ``anterior" but do not go beyond the qualitative reasoning presented and that we repeat for completeness in Section 2.  We shall also see, at the end of section 3, that by choosing a different ``trajectory" for the table, we can get a weaker condition, in fact the same as in Ref. \cite{ref2a}.  Finally, I would like to stress that as we explain in section 4, we have very serious hope to prove ``stability" for tables of which the  feet are on a circle. 

In section 2, we give a sketch of the proof which is based on a fixed point principle, using a continuous  motion of the table. In section 3 we prove, using ``Grandpa$^\prime$s  Mathematics", that such a continuous motion exists, under the sufficient condition that the slope of the ground is less than $\pm 14.4\deg$  with respect to the horizontal plane, and later, $\pm 35.26 \deg$.

\section{
An ÒapproximateÓ proof (sic!)}

The table is supposed to have infinitely narrow feet, and to be invariant under rotations of $90 \deg$ around an axis perpendicular to the table top. The table top is supposed to be sufficiently high or the feet sufficiently long so that there is no risk that, when moved, the top could touch the ground.

For the time being we shall just assume that the ground is continuous and impose stronger restrictions later.

The principle of the proof is extremely simple. We label the 4 feet 1 2 3 4 in their initial positions and we assume that 1 2 3 are \underline{on} the ground. 4 may be on the ground, \underline{above} the ground, or, assuming that the ground can be penetrated, \underline{below}  the ground.

In the first case the problem is solved. Let us take for example the case where 4 is \underline{above} the ground. We invent a continuous motion in which feet 1 2 3 stay on the ground, and which brings 1 to $1^\prime $, coinciding with 2, and 2 to $2^\prime$, coinciding with 3. As for $3^\prime$, it does not coincide with 4, because $3^\prime$ is on the ground while 4  was above. The continuous   motion would be a rotation if the ground were flat. If the ground is smooth, it will be close to a rotation. Of course we are assuming  that this continuous motion exists, and this is what will be proved in the next section.

Now to go from 4 to $3^\prime$ we can perform an exact rotation around the axis defined by 2 3 (equivalent to $1^\prime $ $2^\prime $ ). In this rotation 4 goes towards the ground. The same rotation brings 1 to $4^\prime$; and since 1 was on the ground, $4^\prime$ is below the ground. This means that during the continuous motion  the 4th  foot has gone from above the ground (position 4) to below the ground (position $4^\prime$). (see figure 1). Therefore if the motion is continuous, there is a position where the  4th   foot is on the ground, i.e. the table is in equilibrium.

\begin{figure}
\centerline{\includegraphics[width=.6\textwidth]{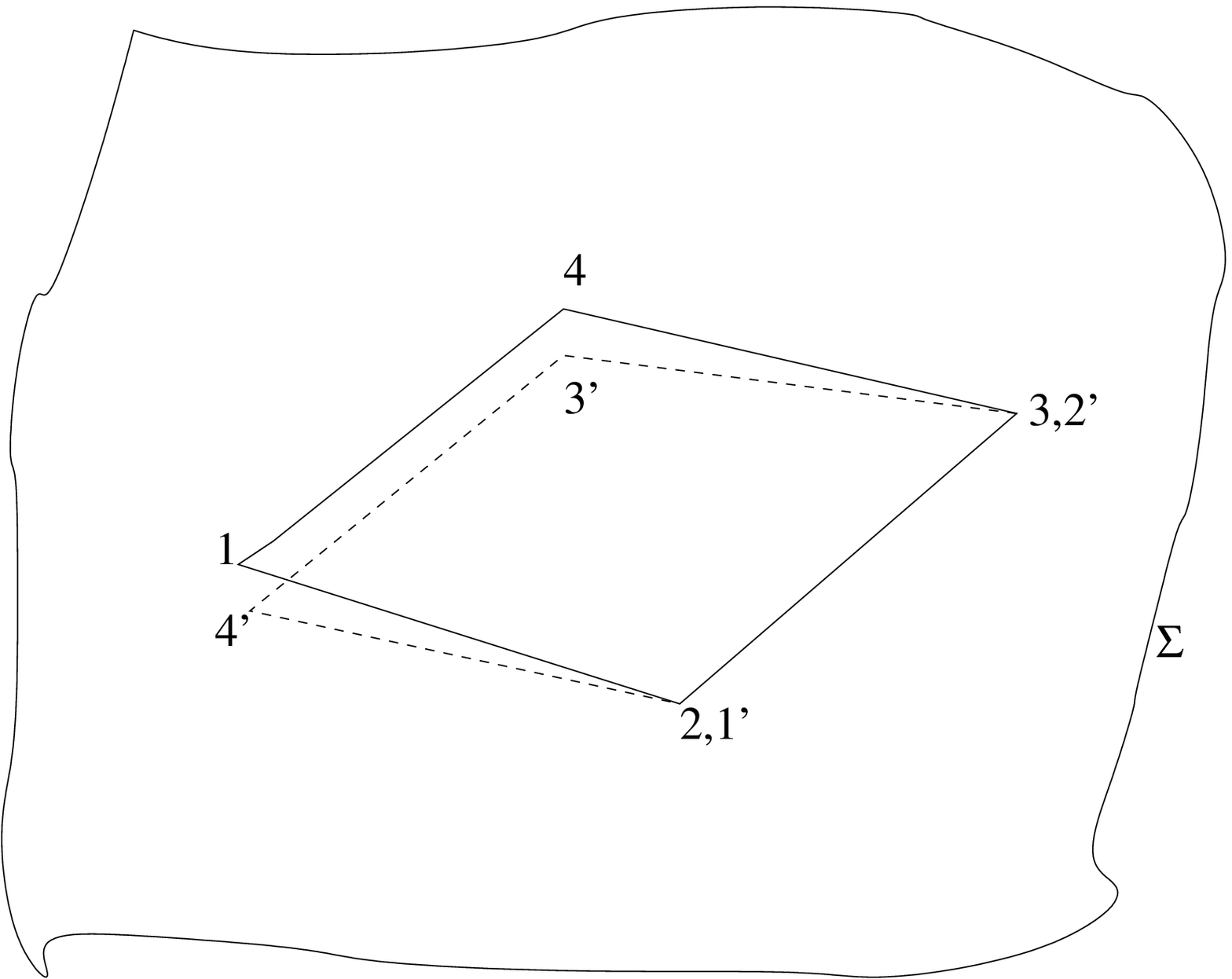}}
\caption{}
\end{figure} 

\section{
A hopefully rigorous proof}

Our problem is to define a continuous motion bringing 1 2 3 4 to $1^\prime 2^\prime 3^\prime 4^\prime$ where 1 2 3 and $1^\prime 2^\prime 3^\prime$ and all intermediate positions are on the ground, and $1^\prime 2^\prime$ coincides with 2 3. This construction is highly non-unique. What we propose is the following:

Unless $3^\prime$ is in the same  plane as 1 2 3, which means that it coincides with 4 and then the problem is solved, the points 1 2 3 $3^\prime$ define a unique sphere S going through them. We call the ground $\sum$ and the intersection of $S$ with $\sum$ the curve $\Gamma$. (Fig. 2).

\begin{figure}
\centerline{\includegraphics[width=.6\textwidth]{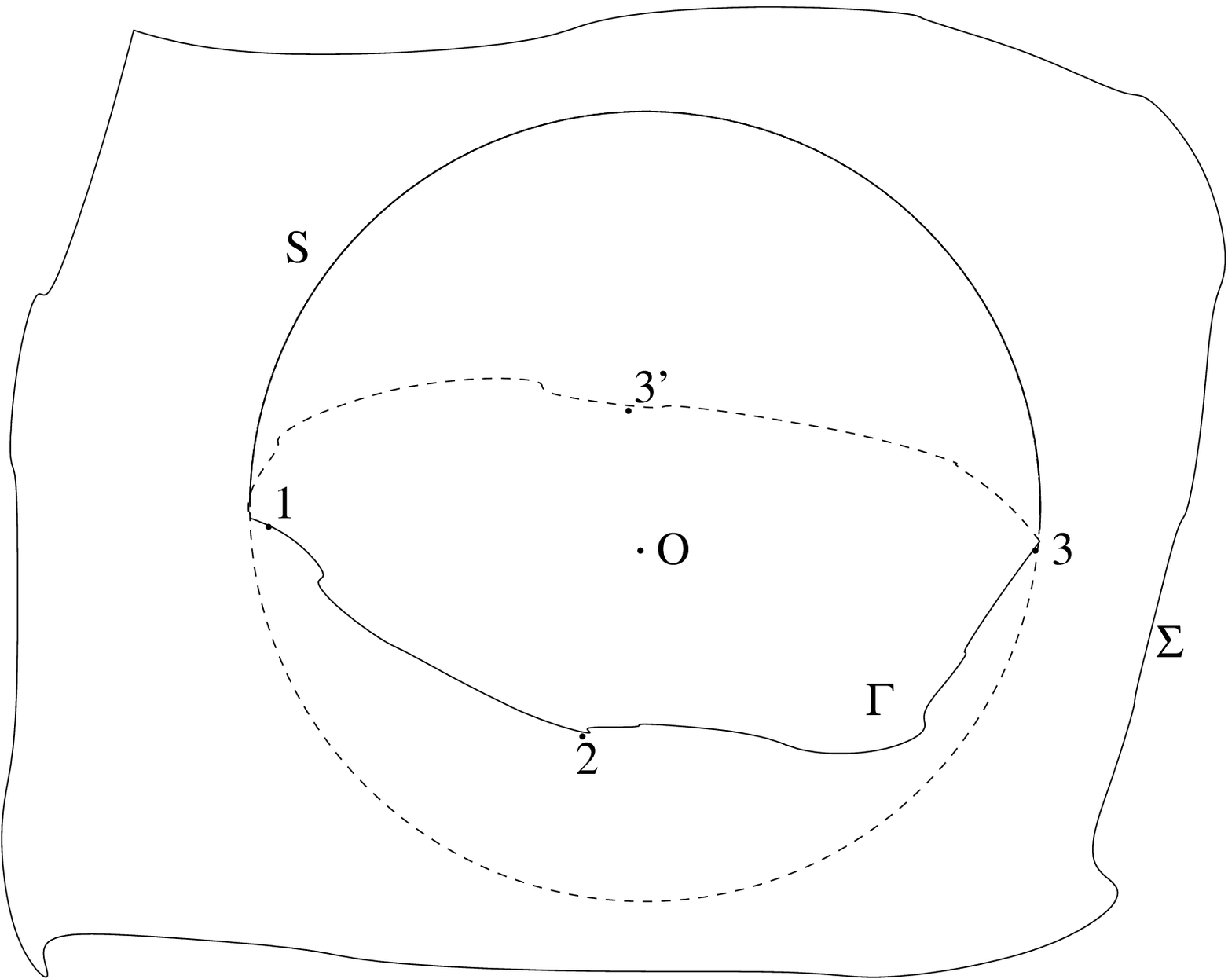}}
\caption{}
\end{figure} 

From now on, we shall assume that, with respect to some reference ``horizontal" plane the slope of the surface does not exceed $\Theta_M$, i.e. given $P$ and $Q$ on $\sum$:

\begin{equation}
\pmatrix {{\vec {PQ}\over  {|PQ|}} .\vec n} \geq \cos \Theta_M ,
\end{equation}
where $\vec{n}$  is the perpendicular to the horizontal plane. We assume further that at any given point there is a tangent plane to the ground.

It turns out to be convenient to impose a condition which guarantees that O, the center of the sphere $SÕ$, is \underline{inside}  the tetrahedron 1 2 3 $3^\prime$.

This tetrahedron has right angles at 2 and 3. If we project it onto the plane perpendicular to 2 3 we get the view shown in Fig. 3.

\begin{figure}
\centerline{\includegraphics[width=.3\textwidth]{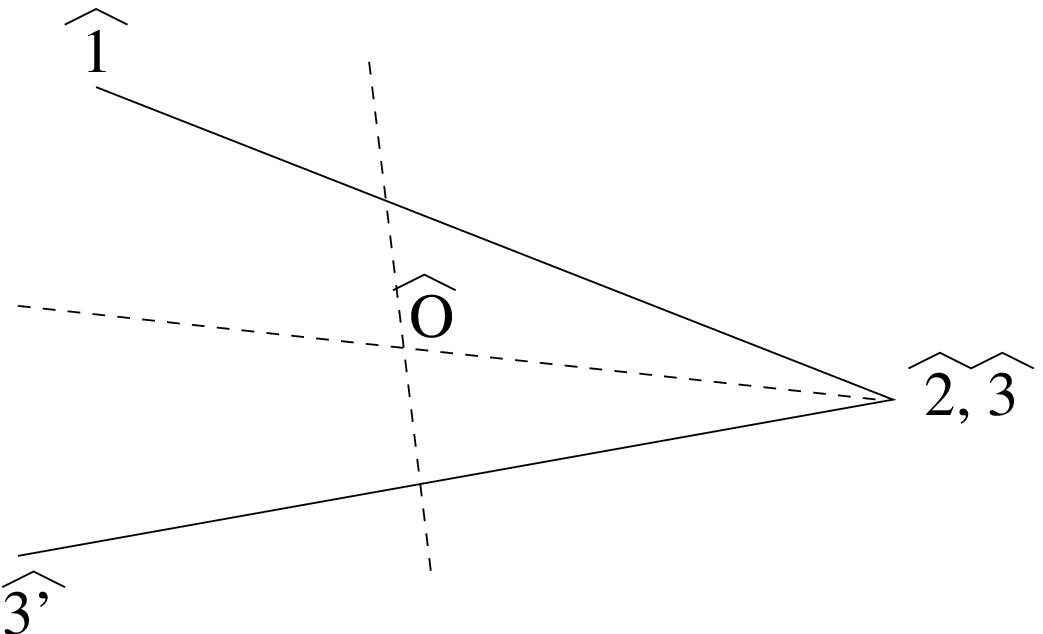}}
\caption{}
\end{figure} 

$\hat{O}$ the image of O, is on the bisector of  $\hat{1} ~~ \hat{23}$ and $\hat {3^\prime} ~~ \hat{23}$, and also on the mediatrice of $\hat {1} ~~\hat{23}$. O will be inside the tetrahedron if the angle $(\hat {23}~~ \hat{1}, \hat{23} ~~\hat{3^\prime})$  or $(\vec{2 1}, \vec {3 3^\prime})$ is less than $\pi \over 2$ in absolute value. On the other hand $\vec {2 3}$ is orthogonal to $\vec {2 1}$ and $\vec {3 3^\prime}$. If we use polar coordinates:

$$\vec {2 1} :  \Theta, \phi$$

\begin{equation}
\vec {2 3} :  \Theta ^\prime, \phi ^\prime\\
\end{equation}

$$\vec {3^\prime 3} :  \Theta ^{\prime\prime}, \phi ^{\prime \prime}~,\\$$
we get 

$$0= \cos (\vec{2 1}. \vec {2 3})=\sin \Theta \sin \Theta ^\prime + \cos \Theta \cos \Theta ^\prime \cos (\phi - \phi ^\prime)$$
\begin{equation}
0= \cos (\vec{2 3}. \vec {3 3Õ})=\sin \Theta ^{\prime \prime}  \sin \Theta ^\prime + \cos \Theta \cos \Theta ^\prime \cos (\phi  ^\prime - \phi ^{\prime \prime})~.\\
\end{equation}

Hence 

$
\vert \cos (\phi - \phi ^\prime) \vert= \vert \tan \Theta \tan \Theta ^\prime \vert \leq (\tan \Theta_M)^2= \sin \alpha\\
$

$
\vert \cos (\phi  ^\prime - \phi ^{\prime \prime}) \vert= \vert \tan \Theta ^\prime \tan  \Theta ^{\prime \prime} \vert <  (\tan \Theta_M)^2= \sin \alpha~.\\
$

So $$\vert \phi -\phi ^\prime \pm {\pi \over 2} \vert \leq \alpha$$

$$\vert \phi  ^\prime -\phi ^{\prime \prime} \pm {\pi \over 2} \vert \leq \alpha~,$$

and \\

$$\vert \phi  -\phi ^{\prime \prime}  modulo~~ \pi \vert < 2 \alpha$$\\

and, since

$\cos (\vec{2 1}. \vec {2 3}^\prime)=\sin \Theta \sin \Theta ^{\prime  \prime} + \cos \Theta \cos \Theta ^{\prime \prime} \cos (\phi - \phi ^{\prime \prime})$\\

\begin{equation}
 \vert \cos (\vec{2 1}. \vec{3 3}^\prime)\vert > (\cos \Theta_M)^2  \cos  2 \alpha -  (\sin \Theta_M)^2,
\end{equation}

\hspace{3cm} if  \hspace{1.5cm}  $\alpha \leq {\pi\over 4}$

Therefore if 

$~~~~~~~~~~~~~~~~~~~~~~~~~~\cos 2 \alpha > (\tan \Theta _M)^2$ , i.e.\\

$$1-2 (\tan \Theta_M)^4 > (\tan \Theta _M)^2$$

Or 
\begin{equation}
 \vert  \tan \Theta _M\vert < {1\over {\sqrt 2}}
\end{equation}
$\cos (\vec{21}, \vec{3 3 ^\prime})$ cannot vanish.

Therefore 0 is inside the tetrahedron  1 2 3 3$^\prime$  if
 \begin{equation}
 \Theta _M < 35.26^{\circ}
 \end{equation}

Then it follows that the points 1 2 3 $3^\prime$ cannot be all on the same side of the horizontal plane going through O. Assume 1 is below this plane and 2 is above. The inclination of $\vec {1 2}$ is less than $\Theta _M$. It is easy to see that this implies that the latitudes of 1 and 2 are less than $2\Theta _M$ in absolute value. Indeed, the latitude of 2 is maximized when 1 is on the equator and 2 is on the opposite meridian. Then the result follows from the theorem of the inscribed angle.

The same applies to 2, 3, $3^\prime$ and to any point of the curve $\Gamma$, because 1 and 2 are on the curve $\Gamma$.

Finally, if  L is the distance between two adjacent feet, the radius $R$ of the sphere $S$ satisfies.

\begin{equation}
 L {1\over {\sqrt 2}} < R < L {{\sqrt 3}\over 2},
\end{equation}
corresponding to the two extreme situations where 1 2 3 $3^\prime$ are close to being in a plane and the one where 1 2, 2 3, and 3 $3^\prime$ are mutually orthogonal.

Let us now find a condition on $\Theta _M$ for ensuring that $\Gamma$ has no double point. We intersect $\Gamma$ by a half plane limited by the vertical line going through O.  If two points of $\Gamma$ are in this half plane, say above the horizontal plane, with latitudes $\Theta _1$ and $\Theta _2$,  the inclination of  the line connecting these two points is ${\pi \over 2} - {{\Theta_1 + \Theta _2}\over {2}}$,  but this should be less than $\Theta _M$. Hence this is impossible if

$$ {\pi \over 2} -2 \Theta_M > \Theta _M, $$i.e.

\begin{equation}
\Theta _M < {\pi \over 6}~~.
\end{equation}

If  $\Theta _M < {\pi \over 6}$ the half plane intersects $\Gamma$ at a single point. This way  it  is possible to define (modulo $2\pi$) the points of $\Gamma$ by their azimuthal angle.

Now we can try to define  the continuous motion going from 1 2 3 to $1^\prime 2^\prime 3^\prime.$ 

The absence of a double point  allows us to order 1 2 3 and $1^\prime 2^\prime 3^\prime$ with  increasing azimuthal  angle and this is true  for any intermediate position that we call 
$1^{\prime \prime} 2^{\prime \prime} 3^{\prime \prime}$. $1^{\prime \prime}$ is determined by the unique intersection of the half plane containing $\vec n$ with $\Gamma$. As we increase  the azimuthal angle of this plane, $1^{\prime \prime}$ moves. $2^{\prime \prime}$ is required to be on $\Gamma$, with a greater azimuthal angle than $1^{\prime \prime}$. $2^{\prime \prime}$ certainly exists because it is at the intersection of a sphere of radius $L$, centered on $1^\prime$, with a sphere of a radius  $>{L\over{\sqrt 2}}$ centered at O passing through $1^{\prime \prime}$, and $\sum$. But is it unique? Even though $\Gamma$ has no double point \underline{this is not obvious}. For this, we shall use continuity, starting from the initial position $1~ 2.$

To make sure we can move the segment of fixed length $1^{\prime \prime} 2^{\prime \prime}$ infinitesimally, it suffices that the tangents to $\Gamma$ at $1^{\prime \prime}$ and $2^{\prime \prime}$ are not orthogonal to $\vec {1 ^{\prime \prime} 2^{\prime \prime}}$.

We call $\vec {1^{\prime \prime} 2^{\prime \prime}}$ $\vec L$, $\vec {O2^{\prime \prime}}$  $\vec R$ and the tangent to $\Gamma$ in $2^{\prime \prime}$ $\vec T$ (Fig. 4). We have, using polar coordinates

\begin{figure}
\centerline{\includegraphics[width=.6\textwidth]{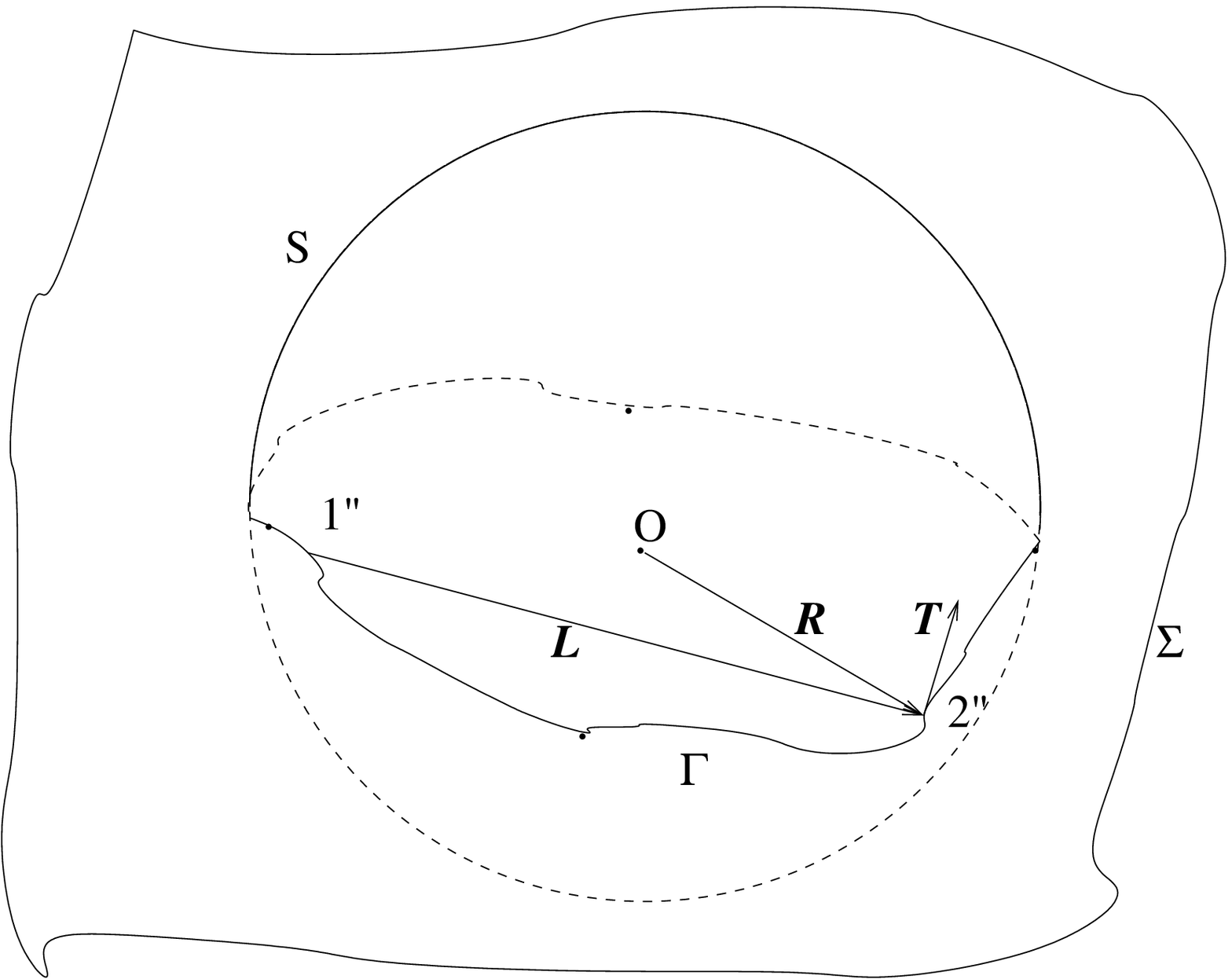}}
\caption{}
\end{figure} 

$$\cos (\vec L, \vec R)= \sin \Theta_L \sin \ \Theta _R + \cos \Theta _L \cos \Theta_R \cos (\phi _L -\phi_R)$$
\begin{equation}
\cos (\vec L, \vec T)= \sin \Theta_L \sin  \Theta _T + \cos \Theta _L \cos \Theta_T \cos (\phi _L -\phi_T)\\
\end{equation}
$$\cos (\vec {R}, \vec T)= \sin \Theta_R \sin \ \Theta _T + \cos \Theta _R \cos \Theta_T \cos (\phi _R -\phi_T)$$
and
\begin{equation}
\vert \Theta_L \vert < \Theta _M, \vert \Theta_T \vert < \Theta_M\\
\end{equation}
$$\vert \Theta_R \vert < 2 \Theta _M
$$

From $$\cos (\vec R, \vec T)=0$$

$$\cos (\vec L, \vec T)=0$$

We now calculate a lower bound on $\cos (\vec L, \vec R)$

We have

$$\vert \cos(\phi_L-\phi_T \vert < (\tan \Theta_M)^2$$
$$\vert \cos(\phi_R-\phi_T \vert < (\tan \Theta_M) \tan 2\Theta_M$$

And then, modulo $\pi,$
$$\vert (\phi_L-\phi_R \vert < \sin^{-1} (\tan \Theta_M)^2 + \sin^{-1} (\tan 2 \Theta_M)^2,$$

$$\vert 
\cos (\phi _L -\phi_R)\vert  > \sqrt {{\cos 3 \Theta_M}\over{(\cos \Theta_M)^5 \cos 2 \Theta_M}} - (\tan \Theta_M)^3 \tan 2\Theta_M$$ 
 and 
 \begin{equation}
 \vert \cos (\vec L, \vec R)\vert > \sqrt {{\cos 2 \Theta_M \cos 3 \Theta_M}\over {(\cos \Theta_M)^3}}-2 {{(\sin \Theta_M)^2}\over {\cos \Theta_M}}
\end{equation}
However, since
$${L\over {\sqrt 3}} < \vert {\vec R} \vert < {L\over {\sqrt 2}}$$

$$\cos (\vec L,\vec R) < {1\over {\sqrt 2}}$$
It is easy to see that if  
\begin{equation}
\Theta_M \leq 14.47^{\circ}
\end{equation}
this is impossible  and therefore, $\vec L$ cannot be orthogonal to $\vec T.$ In the same way, with the same condition, the tangent to $\Gamma$ in $1^{\prime \prime}$ cannot be orthogonal to $\vec L$. The motion of $1^{\prime \prime} 2^{\prime \prime}$ cannot be blocked and, furthermore is monotonous. Condition (12) is probably much too strong.

What remains is to worry about what happens to $3^{\prime \prime}$. During the motion from 1 2 3 to $1^\prime 2^\prime 3^\prime$, $3^{\prime \prime}$ does not necessarily lie on the curve $\Gamma$. We just impose that it stays on the surface $\sum$. We have to show that there is one and only one acceptable intersection of the circle of radius L with center at $2^{\prime \prime}$ and axis 
$1^{\prime \prime} 2^{\prime \prime}$ with $\Gamma$ (in fact there are two intersections, but one of them corresponds to a wrong orientation). Certainly there are intersections. Consider the vertical plane passing through $1^{\prime \prime} 2^{\prime \prime}$ . It contains a perpendicular  to $1^{\prime \prime} 2^{\prime \prime}$ with an inclination superior to $ {{\pi \over 2} -\Theta_M}$, which intersects the circle at two symmetric points.

Since $\Theta _M < 35.26^{\circ}$ one of the points is above $\sum$ and the other below $\sum$. It remains to be shown that the circle intersects $\sum$ in only 2 points, only one of which is acceptable.

\begin{figure}
\centerline{\includegraphics[width=.6\textwidth]{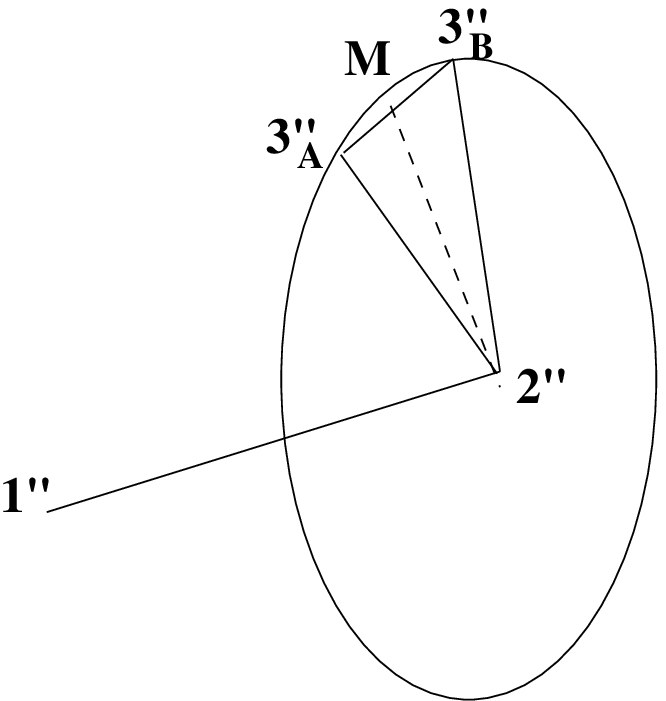}}
\caption{}
\end{figure} 

Assume that the above-mentioned circle has two intersections with $\sum$, as shown in figure 5, $3_{A}^{\prime \prime}$ and $3_{B}^{\prime \prime}$, on the \underline{same} side of the vertical plane passing through $1^{\prime \prime}2^{\prime \prime}$. Call M the middle of $3_{A}^{\prime \prime}3_{B}^{\prime \prime}$. $1^{\prime \prime}2^{\prime \prime}$, $2^{\prime \prime}M$ and $3_{A}^{\prime \prime}3_{B}^{\prime \prime}$ are mutually orthogonal and all have an inclination to the horizontal plane less than $\Theta_M$. This is exactly the situation encountered after equation (2) and hence this is impossible if

\begin{equation}
\tan\Theta_M < {1\over \sqrt2}
\end{equation}

i.e.$$\,\Theta_M < 35.26^{\circ}$$

Under the stronger condition

\begin{equation}
\Theta _M < 14.47^{\circ},
\end{equation}
all obstacles to a continuous monotonous motion from 1 2 3 to $1^\prime  2^\prime 3^\prime$ are lifted. In this motion, the fourth foot has gone from above $\sum$ to below $\sum$, and therefore, there exists an intermediate position where the fourth foot is on the ground.

Now, keeping the same philosophy of moving the table in such a way that the first three feet stay on the ground and that the final position $1^\prime2^\prime$ of the first two feet coincides with 2-3, one can of course invent any continuous movement.  Such a movement, which has the defect of not reducing to a pure rotation for a flat ground , but the advantage of simplicity, consists in (see Fig. 6 for a schematic drawing.) 

\begin{figure}
\centerline{\includegraphics[width=.4\textwidth]{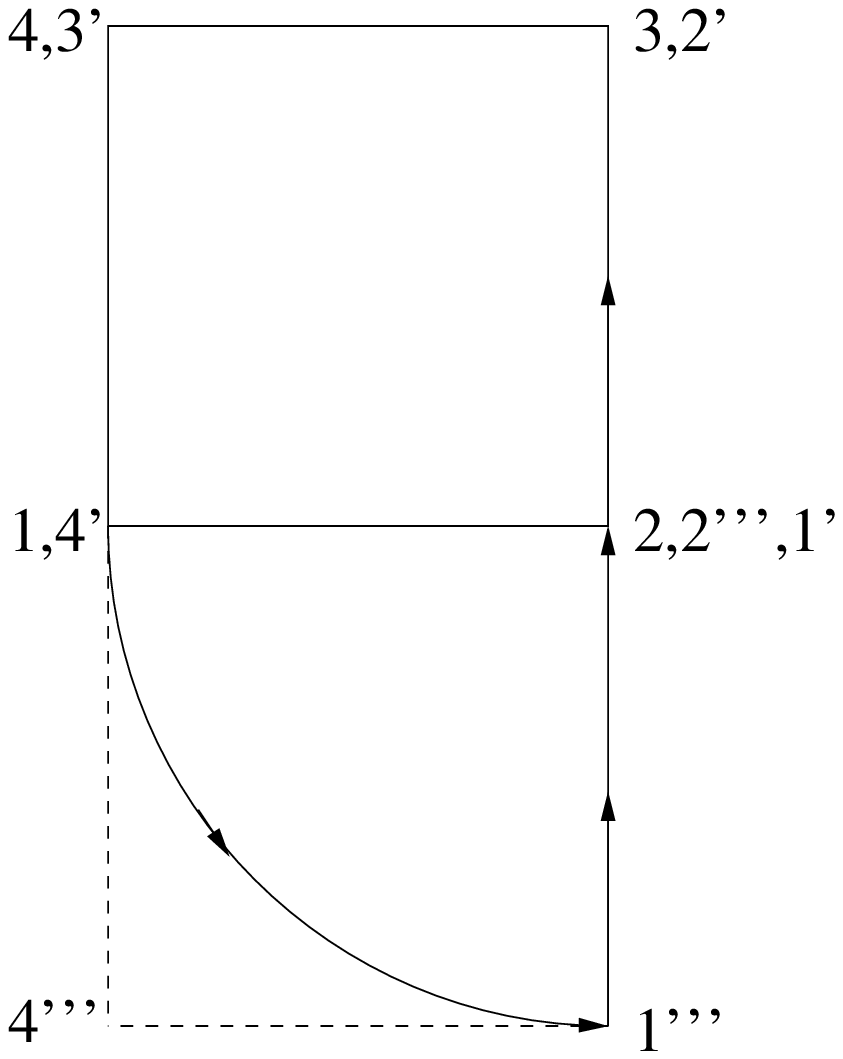}}
\caption{}
\end{figure} 

\begin{itemize}
\item[1)] ``rotating" 12 around 2, keeping 1 on the ground, until 12, which becomes $1^{\prime\prime\prime}2^{\prime\prime\prime}$ is the in the same vertical plane as 23;
\item[2)] Move $1^{\prime\prime\prime}2^{\prime\prime\prime}$ in this vertical plane until it coincides with 23.
\end{itemize}
Of course, during this motion, the third foot is supposed to stay on the ground.

If we look only at the first two feet, we see that during the ``rotation" it is enough to have $\Theta_M = \frac{\pi}{4}$  to make the position of the first foot \underline{unique}, since, under this condition, a half circle centered in 2, containing a vertical axis, will intersect the surface once and only once.  As for the ``translation", it works also under the same condition.

However, we must worry about the third foot.  If we accept a non-monotonous motion of the first two feet, we could probably avoid  imposing any extra conditions.  However, we not only want to have that the 4 feet be \underline{on the ground} but also that the 4 legs be \underline{above} the ground.  This is guaranteed if three mutually orthogonal lines cannot have an inclination less than $\Theta_M$.  This leads to condition (6), which is the same as the one obtained in Ref. \cite{ref2a}. 

 \section{
Concluding remarks  and acknowledgments}

We have proved, under a condition which is certainly too strong, that a perfect square table with four legs always has an equilibrium position. Note that our condition does not require the existence of a tangent plane at all points of $\sum$. There may be conical points. In our proof we have proposed two different motions for the table, the first more natural, the second more efficient, leading to a condition which is in a way optimal if we want the legs of the table to be \underline{above} the ground.  For non square tables, it may be impossible to find an equilibrium position.  
For instance, if the surface $\Sigma$ is a piece of a very large sphere, there will be no local equilibrium position if the 4 feet are not on a circle.  In fact, we believe that if the 4 feet are on a circle, it will be possible to find an equilibrium position on any sufficiently smooth surface, satisfying a condition like (12).  The reason for this belief is this:  put the table horizontal with respect to the reference plane: the 4 feet lie on a circle and they are characterized by the azimuthal angle $\Theta$ of foot  1 with respect to an axis passing through the centre of the circle.  The heights of the feet above (or below) the surface $\Sigma$ are $h_1(\Theta), h_2(\Theta), h_3(\Theta), h_4(\Theta)$.  If we make a complete turn of the table, it is clear that
$$
\int h_1(\Theta)d\Theta = \int h_2(\Theta)d\Theta = \int h_3(\Theta)d\Theta = \int h_4(\Theta)d\Theta
$$
The lines connecting feet 1 and 3 and 2 and 4 intersect in a point $P$, such that $IP = \alpha 13$ and $2P = \beta 24$.  Now, from the previous identity,
$$
\int \left[ \left( (1-\alpha)h_1(\Theta) + \alpha h_3(\Theta)\right) -  \left( (1-\beta)h_2(\Theta) + \beta h_2(\Theta) \right) \right] d \Theta = 0
$$
so that, by the mean value theorem, there is an angle $\bar\Theta$ where the bracket vanishes.  At this point, the opposite intersections $\bar 1, \bar 3$ and $\bar 2,  \bar 4$ of the feet with $\Sigma$ can be connected by straight lines which intersect, so that $\bar 1 \bar 3 \bar 2 \bar 4$ are in a plane.   $\bar 1 \bar 3 \bar 2 \bar 4$ represents an \underline{approximate} equilibrium position of the table.  There are distortions with respect to the exact arrangement of 1234 but these are of higher order with respect to the magnitude of the deviation of $\Sigma$ from an exact plane. In fact these should be two (or an even number) equilibrium position since h($\Theta$) is a function with period 2$\pi$. Whether this can be turned into a rigorous proof or not is not obvious. At present, we can announce that we have an alternative:  either the table has an exact ``equilibrium" position or , for a given small $\Theta_M$ it has infinitely many approximate equilibrium positions, with an error of order $\Theta^3_M$.  An example of a table which is neither square nor rectangular with feet on a circle is a table which is a regular half-hexagon. Such tables in fact exists in some conference rooms at CERN, where they can be used to make convenient arrangements (on flat ground!).  It seems that the method of Ref. \cite{ref2a} works for symmetric trapezo\"\i dal tables, because the diagonals have equal lenghts, but the conditions on $\Theta_M$ depend on the detailed shape of the table.

I thank many collegues, friends, relatives, physicists, mathematicians, or non scientific persons, who have been exposed either to an early version of the proof or to an experiment to``test it", and have encouraged me to publish it. I am grateful to David Dallman for discussions on possible generalizations to non square tables and to Jens Vigen and $\O$yvind $\O$stlund for help in preparing this new version as well as Jean-Marc Richard for the drawings.   I thank A. Krywicki for a useful suggestion and F.J. Dyson for very encouraging comments.  I also thank the Institut des Hautes \'{E}tudes Scientifiques for its past hospitality.

\bibliography{ref}

\end{document}